\documentclass[mathleft
]{an}
\usepackage{graphicx}
\usepackage{times}
\begin{document}

\Pagespan{789}{}
\Yearpublication{2012}%
\Yearsubmission{2012}%
\Month{8}%
\Volume{}%
\Issue{}%

\title{Bulk motion measurements in clusters of galaxies with ATHENA-like missions}

\author{J. Nevalainen\inst{1,2}\fnmsep\thanks{Corresponding author:
  \email{jukka.h.nevalainen@helsinki.fi}\newline}
}
\titlerunning{Bulk motions with ATHENA}
\authorrunning{J. Nevalainen}
\institute{$^1$Tartu Observatory, 61602 T\~oravere, Estonia \\
$^2$Department of Physics, University of Helsinki, Finland
}
\received{ }
\accepted{ }
\publonline{later}


\abstract{The hierarchical formation of clusters of galaxies by accretion of material releases gravitational energy which dissipates into
the intracluster gas. The process heats the material and generates gas turbulence and bulk motions and thus kinetic pressure. 
Mapping the velocity fields of the moving subunits would enable a new diagnostics tool for cluster formation studies and unbiased X-ray mass estimates.
The required spatially resolved high resolution spectroscopy is not currently available. I demonstrate here the feasibility of detecting and mapping 
the velocities of the bulk motions using the Doppler shift of the Fe XXV K$\alpha$ line with the proposed ATHENA satellite.}

\maketitle

\section{Introduction}
During its life cycle, a cluster of galaxies experiences collisions and mergers with other clusters and subunits. Eventually the relaxation processes take place
and the intracluster material approaches hydrostatic and virial equilibrium. Consequently, there will be bulk motions in the cluster material at different distance scales and velocity levels, depending on the magnitude of the event and at which phase we observe the cluster.

Mapping the velocity fields of the bulk motions would open a new tool for studying the dynamics of the clusters.
Comparing the velocity maps with those predicted by cosmological simulations would be useful for the study of the formation of the large scale structure and provide constraints on cosmological parameters.
Combining with the measured galaxy density field, the bulk velocity maps could be used to test whether the fundamental picture of gravitational collapse is correct 
(Branchini et al. 2001; da Costa et al. 1998; Dor\'e et al., 2003).

The bulk motions produce kinetic pressure which can reach a level of ~10\% level of that of the thermal gas even in most relaxed clusters (Lau et al. 2009).
If not accounted for, the kinetic pressure may bias the hydrostatic X-ray mass estimates low by  $\sim$10\% at r$_{500}$.
Thus, it would be important to measure this component in clusters in order to obtain unbiased mass estimates for the cosmological applications. 

In case of recent, strong collisions which happen close to the plane of the sky, the merger shocks can be observed by the X-ray morphology, as in the rare cases
of A520 (Markevitch et al. 2005), A754 (Macario et al. 2011) and A2146 (Russell et al. 2012).
However, in most lines of sight the merger features are hidden by the projection. This allows the possibility of measuring the radial bulk motions via the Doppler shift of the emission lines.

Currently the constrains on radial bulk motions are rather poor due to the lack of spatially resolved high spectral resolution instruments.
The proposed ATHENA mission would have carried such an instrument, an X-ray Microcalorimeter Spectrometer XMS. 
In this paper I will examine the expected quality of bulk motion measurements with a mission approximating the capabilities of 
the proposed ATHENA instruments, i.e. an ATHENA-like mission. 

\section{Bulk motions in clusters}
\subsection{Range of velocities}
It is widely accepted that clusters of galaxies form by merging of smaller structures of matter accreted along large scale structure filaments.
The velocities of the accretion flows are assumed to reach a level of 1000 km s$^{-1}$ (e.g. Frenk et al. 1999). 
A collision of two clusters or protoclusters of similar mass produces a strong merger. The related bulk motion velocities can reach several 
1000 km s$^{-1}$ as in the case of the Bullet cluster (e.g. Markevitch et al. 2002). Simulations of Nagai et al. (2002;2003) indicate that ongoing minor mergers may produce bulk velocities
up to $\sim$1000 km s$^{-1}$. The released gravitational energy is dissipated into the intracluster material which eventually approaches the hydrostatic equilibrium. 
However, simulations indicate that even in the most relaxed clusters there are residual bulk motions present throughout the cluster volume 
at $\sim$100 km s$^{-1}$ level (e.g. Lau et al., 2009; Nagai et al. 2003).

\subsection{Doppler shift}
The radial components of the bulk motions of 100 - 1000 km s$^{-1}$ correspond to 2-20 eV shift of the Fe XXV and XXVI K$\alpha$ emission line centroid energies ($\sim$6 keV). It is challenging to use the currently most powerful X-ray instruments at 6 keV (XMM-Newton/EPIC, Chandra/ACIS and SUZAKU/XIS CCDs) 
for these measurements due to limitations in the energy resolution ($\sim$100 eV). The Gaussian centroid can still be determined with precision better than 100 eV, assuming that the instrument gain is accurately calibrated. This requirement can be relaxed if one considers relatime motions 
between the main cluster and the moving part. In the following I discuss only the statistical precision of the line centroid determination, and 
not the calibration issues nor the uncertainties of the cosmic redshift measurements. 

\subsection{Observational constraints}
Suzaku/XIS instruments have been used to place upper limits for the bulk motion velocities at $\sim$1000 km s$^{-1}$ level in several clusters:
A2319 (Sugawara et al. 2009), Centaurus (Ota et al., 2007), AWM7 (Sato et al. 2008).
A first significant detection of a bulk motion has been achieved with Suzaku for the merging subclump in A2256 (Tamura et al. 2011). The radial velocity difference between the main cluster and the subclump is 1500$\pm$300$\pm$300 km s$^{-1}$ (where the two sets of uncertainties 
refer to statistical and systematical ones, respectively).

\section{ATHENA-like missions}
ATHENA (Advanced Telescope for High ENergy Astrophysics) was one of three L-class (large) missions being considered by the European
Space Agency in the Cosmic Vision 2015-2025 plan. In May 2012 the Jupiter mission Jupiter Icy Moons Explorer (JUICE) was chosen for launch.
However, ESA has committed to continue supporting technology developments for a future large X-ray facility. At the time of writing this paper (Oct 2012) there was an understanding that ESA would shortly appoint a small team from the community to provide input based on the the ATHENA study team activities. 

In this paper I used the responses and background estimates for the X-ray Microcalorimeter spectrometer (XMS)
and a wide field imager (WFI) as reported in the ATHENA assessment report used by the ESA Space Science Advisory Committee (the ATHENA Yellow Book, Barcons et al., 2012).

\subsection{X-ray Microcalorimeter Spectrometer XMS}
The requirement for the energy resolution of XMS is 3 eV at 6 keV. With its 100-1000 times higher effective area at 0.5 keV, compared to those of the current high resolution instruments onboard XMM-Newton and Chandra, and spatial resolution of 10 arcsec,
XMS would enable spatially resolved high resolution spectroscopy.  This would yield a breakthrough in mapping the spectral properties for extended 
sources like clusters. Also, the bandpass extends to 12 keV which allows the measurement of the Fe XXV K $\alpha$ emission line, prominent in clusters of galaxies. 
The downside of XMS is the relatively small FOV (2.3x2.3 arcmin$^2$) which renders the velocity mapping of a whole cluster challenging.

\subsection{Wide Field Imager WFI}
\begin{figure}[h]
\includegraphics[width=80mm]{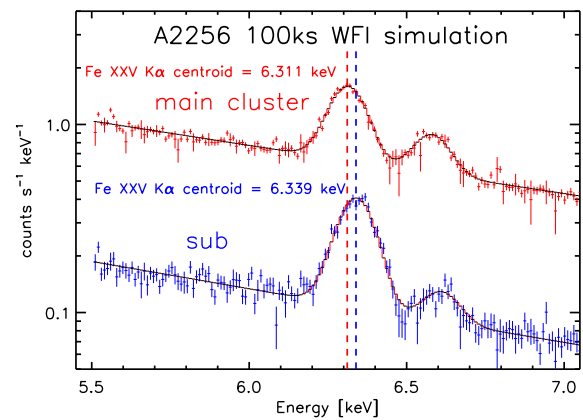}
\caption{Data from a 100ks WFI simulation of the A2256 main cluster (red crosses) and the subclump (blue crosses) together with the input models (solid curves).
The vertical dashed lines highlight the centroids of the Fe XXV K$\alpha$ lines.}
\label{fig2}
\end{figure}

Even though the energy resolution of WFI (150 eV at 6 keV) is not sufficient to resolve the components of Fe K$\alpha$ complex,
it can still determine very precisely the centroid of the Gaussian distribution of the total line emission (see Fig. \ref{fig2}) due to high photon statistics.
This is due to the very large effective area ($\sim$0.5 m$^2$) of the X-ray telescopes onboard ATHENA.
Thus also WFI is  a powerful tool for mapping the cluster velocities.

\section{Improvement with ATHENA-like missions}
I examine here the capability of the proposed ATHENA instruments for measuring the expected bulk motions in clusters.
I used the MEKAL model (Kaastra 1992) in the XSPEC package to model the cluster emission (bremsstrahlung continuum + collisional excitation lines) adopting the metal abundances of Grevesse \& Sauval (1998).

I used the current estimates of the instrument responses (Barcons et al., 2012) to simulate the cluster spectra, including the instrument background and 90\% resolved cosmic X-ray background. Then I fitted the simulated data in the 5.5--7.5 keV band (excluding the channels where the Fe XXVI K$\alpha$ emission is significant)  with a model consisting of a power-law component for the continuum and a Gaussian line for the Fe XXV K$\alpha$ emission. The 1 $\sigma$ statistical uncertainty of the Gaussian centroid then yielded the estimate for the
statistical precision of the velocity measurement, as summarised in Table \ref{velo.tab}.

\subsection{A2256 and the subclump}
In order to obtain a real-life example of the performance of XMS and WFI for measuring the velocities of a nearby minor merger, I used the 
Suzaku results for the A2256 and the subclump (Tamura et al. 2011) to simulate spectra using an exposure time of 100 ks.

\begin{figure}[h]
\includegraphics[width=75mm]{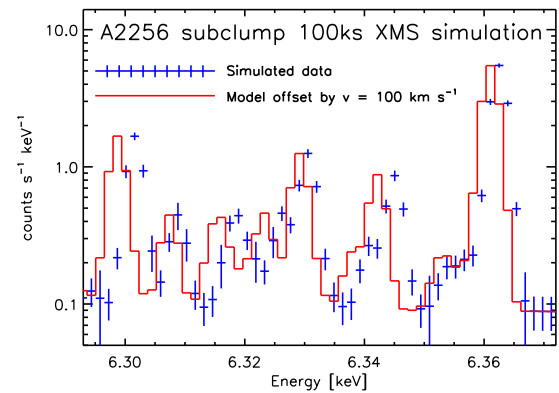}
\caption{Data from a 100ks XMS simulation of the A2256 subclump (blue crosses) together with the input model, shifted by 100 km s$^{-1}$ (red line).}
\label{fig1}
\end{figure}

A fit to the simulated XMS data (see Fig. \ref{fig1}) yielded a statistical uncertainty of the redshift 
($\sigma_z$ $\sim$10$^{-6}$) corresponding to a velocity precision at a level of $\sim$ 1 km s$^{-1}$, i.e. a 1500$\sigma$ detection for the clump motion.
Measurement is very precise because many line features are resolved and each centroid gives weight to $\chi^2$.
Dividing the XMS emission of the subclump into boxes with width of 30 arcsec (i.e. 5$\times$5 pixel map for the full XMS FOV) yielded a statistical precision level of 
$\sim$10 km s$^{-1}$. This level of detail would provide a breakthrough in modelling the dynamics of the merging subclumps. 
Since the flux of A2256 subclump is comparable to that of the central region of similar size in A2256 (Tamura et al., 2011), the above 
calculations yield an approximate estimate of the expected velocity mapping precision level in the bright nearby cluster centres with
XMS. 

A spectral fit to the data from a 100 ks WFI simulation of A2256 main cluster and subclump (see Fig. \ref{fig2}) yielded 
a statistical precision for Fe XXV K$\alpha$ shift corresponding to velocity precision of 60 km s$^{-1}$, i.e. a 
detection at $\sim$25 $\sigma$ level.

\begin{table}
\caption{Velocity precision for a cluster with z = 0.1 and kT = 5 keV with 100ks observation}
\label{velo.tab}
\begin{tabular}{lcccc}\hline
          & \multicolumn{2}{c}{XMS}  &  \multicolumn{2}{c}{WFI}  \\ 
distance $^a$  & $\Delta$r $^b$ & $\sigma_{V}$ &  $\Delta$r & $\sigma_{V}$ \\
(r$_{500}$) &  (r$_{500}$) & (km s$^{-1}$) & (r$_{500}$) & (km s$^{-1}$) \\
\hline
0    & 0.2  & 1   & 0.1 & 100 \\
0    & 0.05 & 10  & --  & -- \\
0.25 & 0.2  & 200 & 0.2 & 200 \\
0.5  & 0.2  & 400 & 0.3 & 800 \\
\hline
\multicolumn{5}{l}{{\bf Notes.} {\footnotesize {\it $^{(a)}$ Distance from the cluster centre. $^{(b)}$ Spatial}}} \\
\multicolumn{5}{l}{ {\footnotesize {\it resolution.} } }\\
\end{tabular}
\end{table}

\subsection{Mapping the whole cluster}
\label{mapping}
In order to derive general conclusions about the performance of ATHENA-like missions for the velocity mapping in clusters, one should simulate 
spectra with a grid of representative values for cluster temperatures, metal abundances and redshifts and realistic exposure times. Also, one should consider a range of luminosities for the moving regions and their velocities and directions of motion. 

In this work I limited the complexity of the above approach by using an exposure time of 100ks for a cluster with kT = 5 keV and z=0.1,  
assuming that all of the emission in the line of sight of a studied region originates from the moving subunit.
I further assumed that the luminosity of the moving subunit at a given distance from the cluster centre can be estimated with a 
a $\beta$ - model for the surface brightness with $\beta$ = 2/3 and r$_{\rm core}$ = 0.1 r$_{500}$. This may be an underestimate 
of the actual signal since the likely enhancement of the emission due to the subunit is not accounted for.
I adopted a bolometric luminosity L$_{bol}$(r$_{500}$) = 7 $\times$ $10^{44}$ erg s$^{-1}$ within r$_{500}$ from L-T relation of Pratt et al. (2009). Using r$_{500}$ - T relation of Vikhlinin et al. (2006) I adopted r$_{500}$ = 10 arcmin.

\subsubsection{XMS}
A single XMS pointing covers only a fraction of the full cluster volume: at z = 0.1 the r$_{500}$ radius for a cluster with kT = 5 keV is $\sim$1 Mpc (Vikhlinin et al., 2006) and covers an area of $\sim$300 arcmin$^2$, i.e. $\sim$50 higher than the FOV of XMS. 
Thus it is not feasible to cover the full r = r$_{500}$ region in nearby clusters with XMS.  
Limiting the mapping into the central r=0.25 r$_{500}$ region would require $\sim$10 pointings which might be feasible for a few clusters. Single XMS pointings at larger radii might still be useful, and I thus calculate estimates for these observations in the following.

I simulated XMS spectra with the above generic cluster (see Section \ref{mapping}) within the full FOV i.e. $\sim$ (0.2 r$_{500}$)$^2$. 
Fitting the simulated spectra yielded a statistical precision of $\sim$200 km s$^{-1}$ at r=0.25 r$_{500}$.
At r=0.5 r$_{500}$ $\sim$90\% of the total signal in the 5.5--7.5 keV band is due to background emission
which degrades the statistical precision of velocity to $\sim$400 km s$^{-1}$ and at r$_{500}$ the data are so noisy due to the background
that reliable velocity measurements cannot be obtained.

If the above cluster was located at z=1.0, a single central XMS pointing would cover the cluster out to $\sim$0.5 r$_{500}$.
The received flux would be similar as that of the z=0.1 cluster at 0.5 r$_{500}$, when using the full 
XMS FOV. Thus, with a central 100ks XMS pointing it is feasible to obtain a single bulk motion measurement at z = 1.0 assuming the full
cluster within 0.5 r$_{500}$ is moving with a radial velocity component higher than 400 km s$^{-1}$.  

\subsubsection{WFI}
Fitting the simulated data using the generic cluster described in Section \ref{mapping} showed that in the centre the bulk velocity can be mapped with angular resolution of $\sim$0.1 r$_{500}$ with statistical precision of $\sim$ 100 km s$^{-1}$.
At r=0.25 r$_{500}$ the lower cluster flux requires a larger extraction region to achieve similar statistical quality as in the centre. 
Using an extraction box with size of ($\sim$0.2 r$_{500}$)$^2$ yields a velocity precision of $\sim$ 200 km s$^{-1}$.
At a distance of 0.5 r$_{500}$ from the centre, the background dominates (as in the case of XMS) and the velocity precision degrades
to $\sim$800 km s$^{-1}$ level when using an extraction box size of $\sim$0.3 r$_{500}$).
Repeating the exercise with a very hot cluster (kT = 10 keV) improves the continuum signal but the Fe XXV emission decreases
and thus the velocity precision does not improve significantly.

\section{Conclusions and discussion}

The simulations showed that for a nearby (z $\le$ 0.1) cluster with kT = 5 keV, using a 100 ks exposure using the instruments proposed for
ATHENA, one could 

\begin{itemize}

\item
measure bulk motions in cluster centres at $\sim$ 0.1 r$_{500}$ spatial scale with statistical velocity 
precision level of $\sim$10 (100) km s$^{-1}$ using XMS (WFI)

\item
map cluster velocities in $\sim$ 10 regions with WFI with a single central pointing up to r = 0.25 r$_{500}$ with $\sim$200 km s$^{-1}$ 
statistical precision with angular resolution decreasing from $\sim$ 0.1 r$_{500}$
to $\sim$ 0.2 r$_{500}$ with increasing radius 

\item
obtain $\sim$200 km s$^{-1}$ precision with XMS up to 0.25 r$_{500}$, but to cover the full cluster within this radius one needs $\sim$ 10 pointings

\item
obtain a velocity precision of $\sim$ 400 km s$^{-1}$ with a single XSM off-axis pointing at 0.5 r$_{500}$ using the emission 
from the full FOV 

\item
obtain a velocity precision of $\sim$ 800 km s$^{-1}$ with WFI at 0.5 r$_{500}$ with a single central pointing

\item
not obtain meaningful velocity measurements at r$_{500}$ due to dominating background

\end{itemize}

The above estimates for the level of statistical precision of velocity measurements with XMS and WFI indicate that an ATHENA-like
mission would enable for the first time the mapping of radial components of the bulk motions due to recent minor and major 
mergers in nearby (z$\le$0.1) clusters of galaxies within the central 0.25 r$_{500}$. Also the residual motions due to past 
mergers can be mapped in these central regions. Single velocity measurements at the distance of 
0.5 r$_{500}$ for a cluster at z=0.1 at the level of $\sim$400 km s$^{-1}$ can be achieved with 100ks off-axis pointings using XMS. 
At higher distances the mapping is more limited due to the background but the average motion within the central 0.5 r$_{500}$ up to z=1.0 can be measured with a central 100ks pointing with XMS, if the radial velocity component is bigger than $\sim$400 km s$^{-1}$.

These measurements would open a new tool to study cluster dynamics. This would be complementary to the current analyses of 
intracluster shocks and sloshing and thus give a better handle on the 3-dimensional dynamics of clusters. This would yield a breakthrough 
in modelling the cluster physics and its implications to cosmology. 

The kinetic Sunyaev Zeldovic (kSZ) effect as measured by WMAP (Kashlinsky et al., 2008; Osborne et al., 2012) has been used to constrain 
bulk motions of clusters. However these results have been controversial due to relatively weak kSZ signal compared to the 
thermal one. Thus, one must combine a large sample of clusters and assume a similar motion for all the clusters in the sample.
Mak et al. (2011) calculate that a sample of 400 clusters would allow Planck to detect a coherent 500 km s$^{-1}$ bulk motion.
Clearly ATHENA would make improvement here, being able to measure the velocities in individual clusters with better precision and 
spatial resolution.




\begin{thebibliography}{}
  \bibitem{} Barcons, X., Barret, D., Decourchelle, A. et al., arXiv:1207.2745
  \bibitem{} Branchini, E., Freudling, W., Da Costa L.N et al.: 2001, MNRAS 326, 1191
  \bibitem{} da Costa, L.N.,  Nusser, A.,  Freudling, W. et al.: 1998, MNRAS, 299, 425
  \bibitem{} Doré, O., Knox, L., \& Peel, A.: 2003, ApJ, 585L
  \bibitem{} Frenk, C. S., White, S. D. M., Bode, P. et al.: 1999, ApJ, 525, 554
  \bibitem{} Grevesse, N., \& Sauval, A., 1998, SSRv, 85, 161
  \bibitem{} Kaastra, J. 1992, An X-Ray Spectral Code for Optically Thin Plasmas (Internal
             SRON-Leiden Report, updated version 2.0)
  \bibitem{} Kashlinsky, A., Atrio-Barandela, F., Kocevski, D., et al., 2008, ApJL, 686, 49 
  \bibitem{} Lau, E. T., Kravtsov, A. V. \&  Nagai, D.: 2009, ApJ, 705, 1129
  \bibitem{} Macario, G.,  Markevitch, M.,  Giacintucci, S. et al.:	2011, ApJ, 728, 82
  \bibitem{} Mak, D.,  Pierpaoli, E. \& Osborne, S., 2011, ApJ, 736, 116 
  \bibitem{} Markevitch, M.,  Gonzalez, A. H., David, L. et al.: 2002, ApJ, 567, L27
  \bibitem{} Markevitch, M., Govoni, F.,  Brunetti, G. et al., 2005, ApJ, 627, 733
  \bibitem{} Nagai, D.,  Kravtsov, A. V. \& Kosowsky, A.: 2003, ApJ, 587, 524
  \bibitem{} Osborne, S., Mak, D., Church, S. et al., 2012, submitted arXiv:1011.2781v2
  \bibitem{} Ota, N., Fukazawa, Y.,  Fabian, A. C., et al.: 2007, PASJ, 59, 351
  \bibitem{} Pratt, G. W., Croston, J. H.,  Arnaud, M., et al.: 2009, A\&A, 498, 361
  \bibitem{} Russell, H. R., McNamara, B. R., Sanders, J. S. et al.: 2012, MNRAS, 423, 236  
  \bibitem{} Sato, K., Matsushita, K., Ishisaki, Y. et al.: 2008, PASJ, 60, 333
  \bibitem{} Sugawara, C., Takizawa, M. \& Nakazawa, K.: 2009, PASJ, 61, 1293
  \bibitem{} Sun, M., Murray, S. S., Markevitch, M. et al.: 2002, ApJ 565, 867
  \bibitem{} Tamura, T., Hayashida, K., Ueda, S. et al.: 2011, PASJ, 63S1009
  \bibitem{} Vikhlinin, A., Kravtsov, A.,  Forman, W. et al.,  2006, ApJ, 640, 691
 \end{thebibliography}
\end{document}